\documentclass{article}

\PassOptionsToPackage{numbers, compress}{natbib}
\usepackage[preprint]{neurips_2026}

\usepackage[utf8]{inputenc} 
\usepackage[T1]{fontenc}    
\usepackage{hyperref}       
\usepackage{url}            
\usepackage{booktabs}       
\usepackage{amsfonts}       
\usepackage{nicefrac}       
\usepackage{microtype}      
\usepackage{xcolor}         

\usepackage{graphicx}
\usepackage{multirow}
\usepackage{pdfpages}
\usepackage{enumitem}
\usepackage{xspace}
\usepackage{amsfonts, amsmath, amssymb}
\usepackage{xcolor}
\usepackage{multirow}
\usepackage{graphicx}
\usepackage{hyperref}

\usepackage{wrapfig}
\usepackage{graphicx}
\usepackage{hyperref}
\usepackage{multirow}
\usepackage{graphicx}
\usepackage{tabularx}
\usepackage{array}
\usepackage{url}
\usepackage{xcolor}
\usepackage{epsfig}
\usepackage{adjustbox}
\usepackage{amsfonts, amsmath, amssymb}
\usepackage{booktabs} 
\usepackage{comment}
\usepackage{caption, subcaption}
\usepackage{textcomp}
\usepackage{relsize}
\usepackage{stmaryrd}
\usepackage{bbm}
\usepackage{rotating}
\usepackage{helvet}
\usepackage{cleveref}
\usepackage{xspace}
\usepackage{enumitem}
\usepackage{epigraph}
\usepackage{mdframed}
\usepackage{makecell}
\usepackage{tcolorbox}
\usepackage{nicematrix}
\usepackage{tabu}
\usepackage{colortbl}
\usepackage{xcolor}
\usepackage[table]{xcolor}
\PassOptionsToPackage{table}{xcolor}
\newcolumntype{L}{>{\raggedright\arraybackslash}X}
\newcolumntype{C}{>{\centering\arraybackslash}p{0.105\columnwidth}}

\title{Will It Go Viral? Grounding Micro-Video \\Popularity Prediction on the Open Web}

\author{%
  Ryang Heo \quad Dongha Lee\thanks{\; Corresponding author} \\
  Department of Artificial Intelligence, Yonsei University \\
  \texttt{\{ryang1119, donalee\}@yonsei.ac.kr} \\
}

\newcommand{\dataset}{\textsc{WebShorts}\xspace}
\newcommand{\proposed}{\textsc{Shorts-Cast}\xspace}

\begin{document}

\maketitle

\begin{abstract}
Micro-video popularity prediction (MVPP) forecasts the popularity a newly uploaded short-form video will attract within a fixed number of days after upload.
This task supports downstream applications in recommendation, advertising, and creator analytics, yet the problem is hard since virality depends on external trends rather than video content alone.
Prior MVPP methods incorporate context by retrieving similar videos from platform-internal corpora, however historical neighbors cannot reveal whether a topic is currently trending, controversial, or already saturated across the open web.
To this end, we reformulate MVPP as \textit{open-web grounded prediction} and introduce \dataset, the first micro-video dataset that couples 14K videos with real-time open-web context collected at upload time, alongside daily view counts tracked over 7 days. 
The context for each video is organized as a structured \textit{evidence-card} that captures the external attention landscape along three complementary web-context dimensions.
We further propose \proposed, a framework that generates dimension-wise rationales from the evidence-card to guide popularity regression, then adapts at deployment by selectively updating the context-to-popularity mapping when delayed labels reveal genuine trend shifts.
In our experiments, \proposed consistently outperforms content-only, video corpus retrieval-augmented, and online adaptation baselines under both offline and delayed-label online protocols, confirming that structured web context and trend-aware adaptation are jointly necessary for popularity forecasting under realistic deployment constraints in fast-evolving short-form video ecosystems.
\href{https://github.com/ryang1119/Shorts-cast}{[CODE]}
\end{abstract}

\section{Introduction}
\label{sec:intro}
Micro-videos have become a dominant form of online content on platforms such as YouTube Shorts, TikTok, and Kuaishou.
Predicting their popularity~\cite{chen2016microtellsmacro, wu2019smp, ye2025mvp} is important for recommendation, marketing, and creator support~\cite{xie2020mmved, zhong2024mmra, cheng2025seeing}, yet difficult in practice: some videos surge immediately after upload, while others gain traction days later as external events, memes, or public discourse shift attention toward their topic.
The popularity of a micro-video is therefore not determined by its content alone but is shaped by the broader context surrounding its topic at discovery time~\cite{hessel2017cats, rizoiu2017hip}.

\begin{figure}
    \centering
    \includegraphics[width=0.99\linewidth]{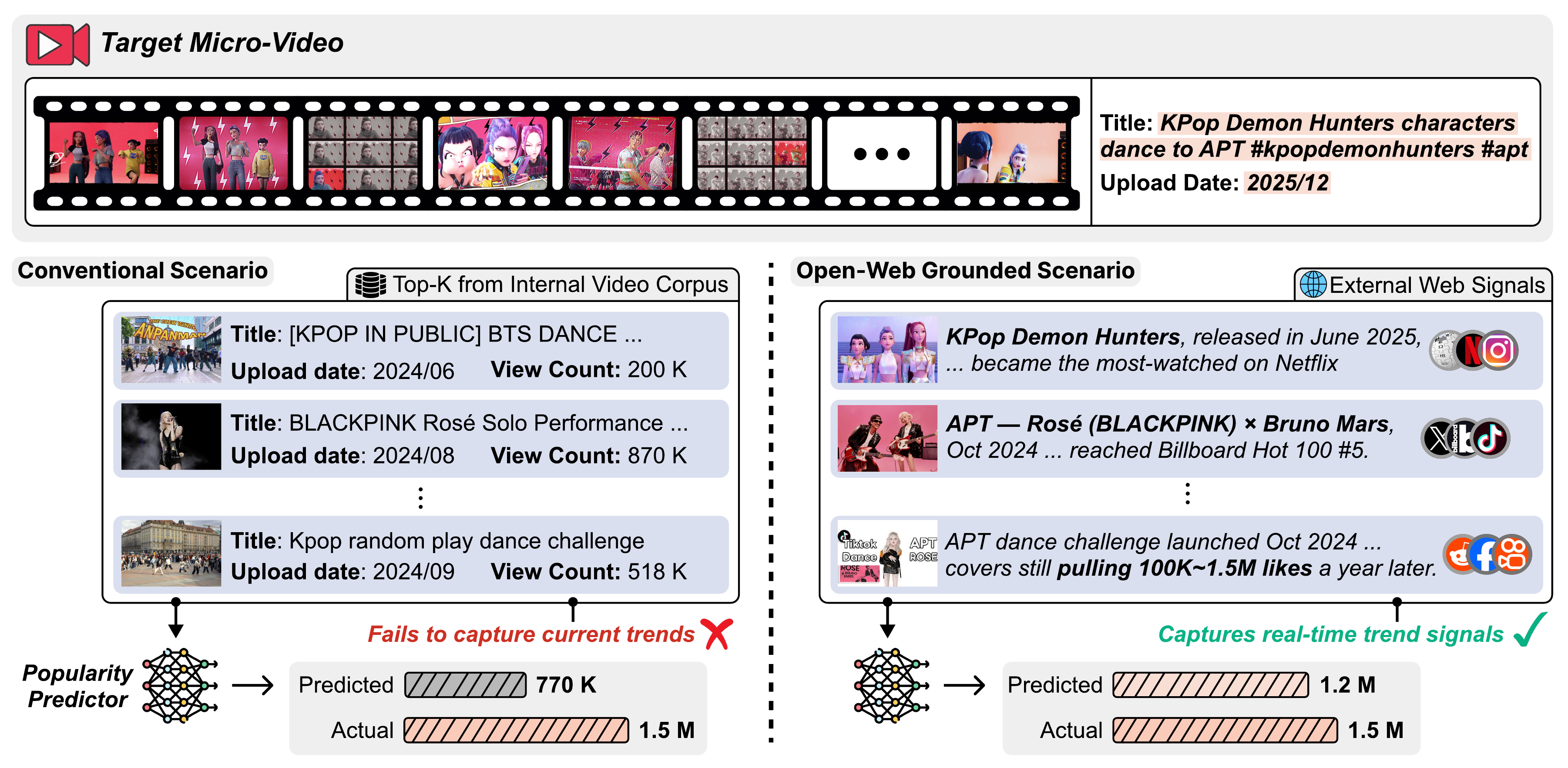}
    \caption{Conventional scenario \textbf{(Left)} retrieves similar videos from a static in-platform corpus, missing the external trends behind virality. In contrast, the open-web grounded scenario \textbf{(Right)} captures real-time web signals, closely anticipating the viral outcome.}
    \label{fig:intro_fig}
    \vspace{-0.7cm}
\end{figure}

Recent micro-video popularity prediction (MVPP) research has attempted to incorporate such context by retrieving similar videos from an internal video corpus and using their features and popularity labels as auxiliary prediction signals~\cite{zhong2024mmra, cheng2025seeing, cheng2024ragtrans, chen2025echoes, cheng2025incontext, xu2025skapp}.
This approach improves over content-only prediction, but the context it provides is limited to historical in-platform videos.
As shown in Figure~\ref{fig:intro_fig}, such videos can reflect past popularity patterns, but they cannot reveal whether the topic currently commands public attention, or competes with an already saturated content pool on the open web.
This motivates us to reformulate MVPP as \textbf{open-web grounded prediction}, where future popularity is predicted from external web context.

However, extending MVPP from internal-corpus retrieval to open-web grounded prediction is not straightforward, as this shift surfaces two challenges that prior formulations leave unaddressed.
\textbf{(1) Prediction-time web context alignment}: Since MVPP forecasts future popularity at prediction time, the external web context used for prediction must reflect what was observable at that moment.
Existing MVPP datasets~\cite{chen2016microtellsmacro, cheng2025seeing, ni2023content, xu2025smtpd} provide video content, metadata, and popularity outcomes (i.e., view count), but do not preserve the web context surrounding each video at prediction time.
Retrospectively collecting such context is unreliable: search results and public discourse shift continuously, and later snapshots risk encoding signals about the video's realized popularity, creating label leakage.
Open-web grounded MVPP therefore requires prospective collection of external web context aligned to each video's prediction time.
\textbf{(2) Trend-adaptive context-to-popularity mapping}:
Even with properly aligned web context at every prediction, the mapping from context to popularity drifts as the surrounding trend landscape shifts.
This is especially acute in short-form platforms, where last week's viral threshold becomes this week's baseline and emerging content formats introduce patterns absent from the training distribution~\cite{chen2025echoes, gong2022real}.
A model trained offline on a fixed snapshot of videos, web context, and popularity labels cannot track these shifts.
The predictor must therefore keep learning how the current trend landscape shapes the relationship between web context and video popularity.

To tackle these challenges, we introduce \dataset, the first MVPP dataset that pairs each micro-video with open-web context collected in real time as videos are uploaded and tracked.
\dataset contains 14K video instances with daily view-count recordings over 7 days after upload, each annotated with a structured \textit{evidence-card} that captures the external attention landscape surrounding each video's topic into three dimensions: (i) topic \& entity context, (ii) public discourse, and (iii) related content activity.
Evidence-cards are collected at three upload-relative observation snapshots, capturing how the web context surrounding a video topic evolves during the early observation window.

Building upon \dataset, we present \proposed, a framework for open-web grounded MVPP that learns to reason over structured web context and adapts to shifting trends at deployment time.
\proposed first trains a predictor that generates dimension-wise rationales from the evidence-card, articulating how each evidence dimension contributes to the video's expected popularity before producing the scalar prediction.
Once deployed on a live stream, \proposed adapts to changing trends through \textit{Growth-Conditioned Drift} filtering.
This strategy uses view-growth patterns to distinguish genuine trend-driven prediction failures from routine noise, and triggers lightweight updates to keep the context-to-popularity mapping current while preserving the base predictor.


We summarize our main contributions as follows:
\begin{itemize}[leftmargin=*,topsep=2pt,itemsep=2pt,parsep=0pt]
    \item We formulate MVPP as open-web grounded-prediction, a paradigm that forecasts future popularity from external web context, moving beyond internal-corpus retrieval of historical platform videos.
    \item We introduce \dataset, the first MVPP dataset that tracks each micro-video with daily popularity snapshots and grounds it in real-time  open-web context, captured as evidence-cards with three complementary dimensions.
    \item We propose \proposed, a framework that generates dimension-wise rationales from the evidence-card to guide popularity regression, and adapts the learned context-to-popularity mapping at deployment by detecting trend-driven failures from view-growth patterns.
\end{itemize}

\section{Related Work}
\label{sec:relwork}

\paragraph{Micro-Video Popularity Prediction}
Micro-video popularity prediction (MVPP) aims to forecast the engagement that a newly uploaded short-form video will accumulate within a future time horizon~\cite{chen2016microtellsmacro, wu2019smp, ye2025mvp, wu2023smp}.
Early works relied on hand-crafted signals from user profiles, visual content, and posting time, fed to classical regressors~\cite{khosla2014what, jing2018lowrank, chen2019social, lai2020hyfea}.
Content-centric methods then replaced these manual features with deep multi-modal fusion and pre-trained vision-language models (VLMs), but still primarily modeled cues within the target video itself~\cite{xie2021micro, cheung2022crossmodal, zhang2022multimodal, radford2021learning, du2023multi, hu2024dual, tu2024higher}.
A more recent and prominent line moves beyond isolated target-video modeling by augmenting prediction with retrieved historical videos, in-context prompts, graph-structured support sets, or modality-recovery signals as auxiliary context~\cite{zhong2024mmra, cheng2025seeing, cheng2024ragtrans, chen2025echoes, cheng2025incontext, xu2025skapp}.
All these methods, however, remain structurally blind to whether the target topic is currently amplified or saturated across the open web, the gap our formulation addresses.
We therefore formulate MVPP as open-web grounded prediction, shifting the task toward forecasting future popularity from temporally aligned external web context.


\paragraph{Online Adaptation}
Online adaptation has recently been studied across vision-language recognition~\cite{karmanov2024efficient, 
zanella2025realistic, cao2025noisy, han2026dota}, time-series forecasting~\cite{lee2025lightweight, zhao2025proactive, lau2025fast}, recommendation system~\cite{lee2024continual, liao2025mitigating}, and streaming data analysis~\cite{wan2024online}, where deployed models must cope with non-stationary inputs and delayed or evolving feedback.
Across these domains, the key objective is to adapt to changing data distributions without full retraining.
Existing methods typically address this challenge through lightweight adaptation mechanisms, including low-rank updates~\cite{liang2024inflora, wei2025onlinelora, lianggated}, prompt tuning~\cite{zhang2025hierarchical, zhangdpcore}, and cache-based memories~\cite{karmanov2024efficient}, while keeping most backbone parameters fixed.
For open-web grounded MVPP, this need is particularly important: even when fresh web context is available at inference time, the learned mapping from context to popularity can become outdated as topics, public discourse, platform exposure, and competing content evolve in live deployment.

\section{\textsc{WebShorts}: Open-Web Grounded Dataset for MVPP}
\label{sec:dataset}
\begin{figure}
    \centering
    \includegraphics[width=0.95\linewidth]{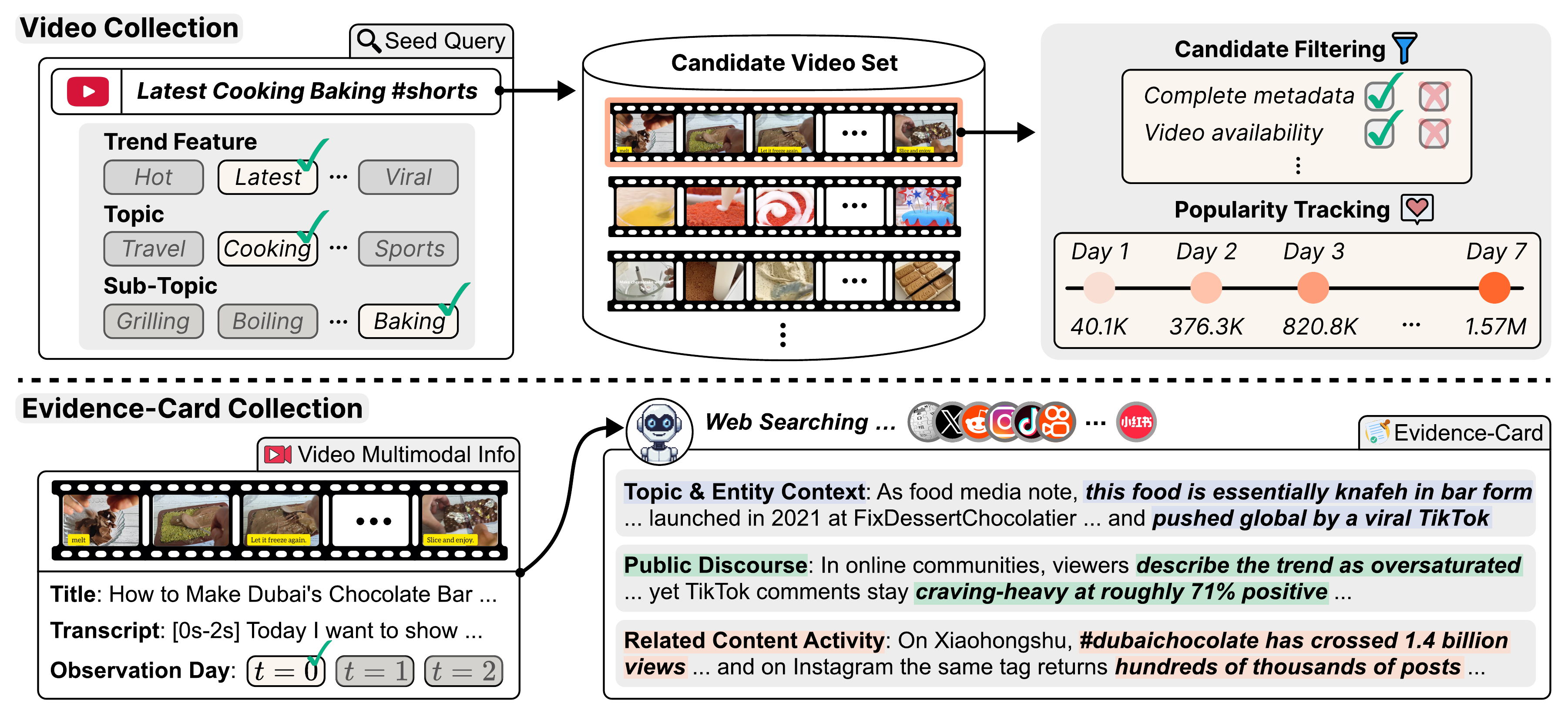}
    \caption{The overview of our \dataset construction pipeline.}
    \label{fig:dataset}
\end{figure}

We present \dataset, a micro-video dataset designed for open-web grounded popularity prediction. 
Unlike conventional datasets that pair each video only with platform-internal metadata and popularity labels, \dataset augments each target video with time-aligned open-web context through a structured \textit{evidence-card}. 
Figure~\ref{fig:dataset} illustrates the overall construction pipeline.



\subsection{Task Formulation}
Recent approaches in MVPP~\cite{zhong2024mmra, cheng2025seeing, chen2025echoes} typically aim to predict the cumulative popularity of a given micro-video, that is, its post-release popularity trend.
In this work, we formulate the task as predicting the popularity of a micro-video 7 days after upload.
Formally, let $V = \{V_1, \ldots, V_N\}$ denote the set of micro-videos, where $N$ is the total number of micro-videos.
Given a micro-video $V_i$, the goal is to predict its view count at day $T$ after upload.
Here, popularity is measured as the cumulative view count, and following prior works~\cite{cheng2024ragtrans, xu2025smtpd, xu2025forecastingbuzz}, we define the target popularity $y_i$ of $V_i$ on a log scale of the raw view count $v_i(T)$ as $y_i = \log_2\big(v_i(T) + 1\big)$.

\subsection{Video Collection}
To ground our dataset in one of the largest and most actively used short-form video platforms, we collect all videos in \dataset from YouTube Shorts.
YouTube Shorts is a short-form video service that hosts videos with a duration of no longer than three minutes.
Following prior work~\cite{xu2025smtpd}, we construct a temporally aligned dataset by collecting newly available videos and tracking their popularity after upload.
To this end, we utilize the YouTube Data API to retrieve video candidates and their associated metadata.

\paragraph{Candidate retrieval}
To support diversity in video topics and categories, we adopt the hierarchical topic categorization from prior video datasets~\cite{qiu2024mmsum, lee2025hippovideo}, comprising 17 main topics and 10 sub-topics per main topic~\cite{zhou2018towards, miech2019howto100m}.
We further introduce a trend feature axis (e.g., Hot, Latest, Viral) with 10 variants, and construct seed queries by combining all three axes into templates such as ``\textit{Latest Cooking Baking \#shorts}'', yielding 1,700 seed queries that capture the breadth of content on YouTube.
Details on the full query composition are provided in Appendix~\ref{subsec:appendix_video_collection}.

\paragraph{Candidate filtering} 
After retrieving candidate videos, we apply filtering criteria to ensure data quality.
Candidates without complete metadata (e.g., title, upload time, or channel information) or whose videos are no longer accessible are excluded.
We further adopt Whisper-large-v3~\cite{radford2023robust} to generate automatic speech recognition (ASR) transcripts, which are used to identify English-language videos and filter out candidates that do not satisfy the target language condition.

\paragraph{Popularity tracking} 
For all selected videos, we continuously fetch view counts for seven days from the upload time. The initial observation immediately after upload is stored as Day 0, and we then collect daily snapshots of cumulative view counts from Day 1 to Day 7. 
Since videos are uploaded at different absolute times, we use a relative day index based on each video's own upload time, where each observation day is defined by adding 24-hour intervals to the original upload time. 
This design ensures that the popularity measurements are aligned by the same post-upload horizon across videos.

\subsection{Evidence-Card Collection}
\paragraph{Evidence-card design}
The evidence-card is the core component that distinguishes \dataset from existing MVPP datasets.
Rather than attaching raw web search results, we construct a structured evidence-card $E_i$ for each video $V_i$ that organizes the open-web context surrounding the video topic into three complementary dimensions:
(D1) \textit{Topic \& entity context}: factual background, relevant events, entities, and timelines;
(D2) \textit{Public discourse}: public reactions, sentiment patterns, controversies, and community discussion; and
(D3) \textit{Related content activity}: similar or competing content across platforms that may dilute or amplify the target video's attention.
These three dimensions cover the distinct mechanisms through which external context shapes popularity: D1 captures \textit{what} the topic is and how established it is, D2 captures \textit{how} the public perceives it, and D3 captures \textit{how much} competing content already occupies the same attention space.

\paragraph{Evidence-card construction}
With the advent of search-augmented large language models (LLMs)~\cite{perplexity2024, openai2024chatgptsearch, comanici2025gemini}, it has become feasible to search the open web, retrieve information from diverse sources, and organize it into structured forms~\cite{li2025search, miroyan2026search, liang2025videobrowser, kim2026agenticshop}.
Building on this capability, we construct an evidence-card for each video using a search-augmented LLM.
Specifically, we provide grok-4.1-fast-reasoning~\cite{grok2025} with the multi-modal content $X_i$ of each video as interleaved image-text input, including the title, four key frames sampled at the 0\%, 25\%, 50\%, and 75\% positions, and the ASR transcript, where the 0\% frame corresponds to the thumbnail.
We then generate an evidence-card $E_i^{(t)} = \texttt{LLM}_{\text{search}}(X_i, t)$ aligned with observation day $t$.
Motivated by the importance of temporal alignment and early popularity in social media popularity prediction~\cite{xu2025smtpd}, we instantiate $t$ as the first three relative observation days since upload, i.e., $t \in \{0,1,2\}$, which captures early contextual signals while remaining sufficiently distant from the day-7 popularity outcome.
To prevent platform leakage, we exclude the YouTube domain from the search scope via \texttt{excluded\_domains}, ensuring that the target video's view count or platform-internal popularity signals cannot enter the evidence-card.

\subsection{Dataset Analyses}
\begin{figure}
    \centering
    \includegraphics[width=0.99\linewidth]{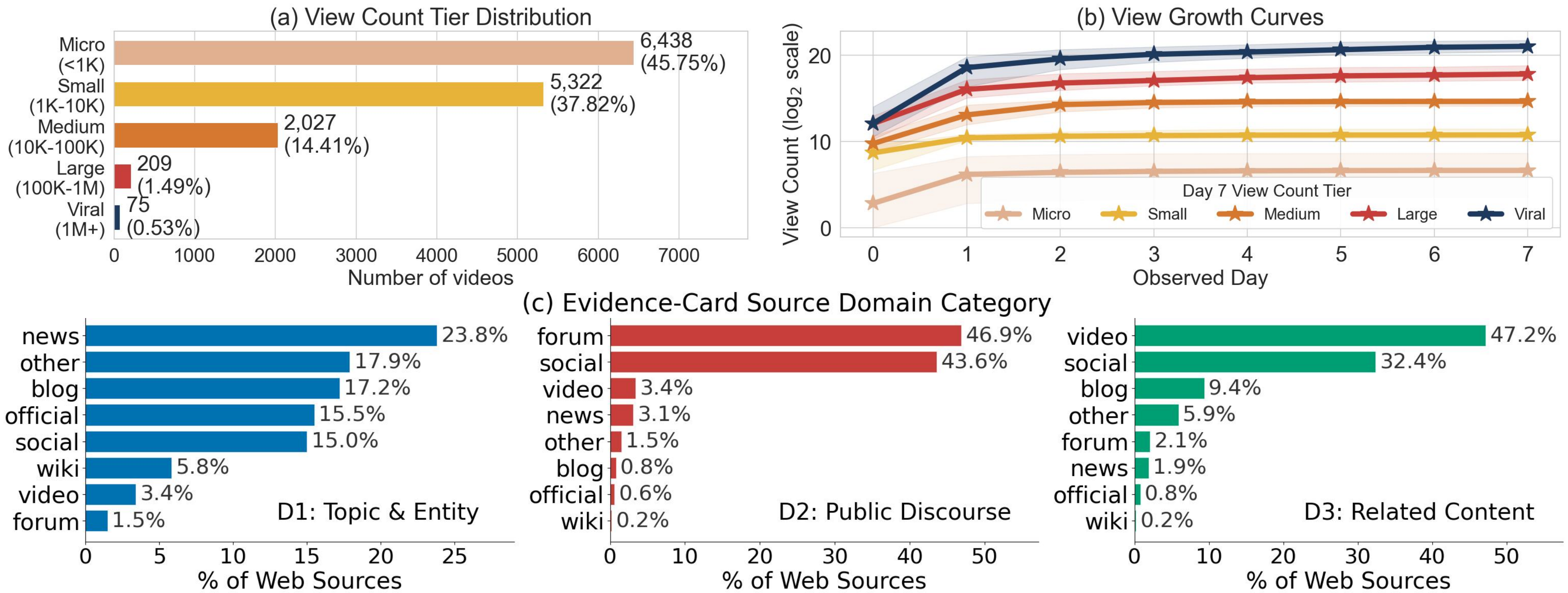}
    \caption{\dataset statistics: (a) day-7 view count distribution by popularity tier, (b) per-tier view growth curves, and (c) web source distribution across evidence-card dimensions.}
    \label{fig:stats_dataset}
\end{figure}

\paragraph{Overall statistics}
\dataset contains 14K YouTube Shorts collected over a one-week upload window.
As shown in Figure~\ref{fig:stats_dataset} (a, b), the day-7 popularity distribution is heavily long-tailed, following the pattern observed in prior social media datasets~\cite{xu2025smtpd}, with Micro ($<$1K) and Small (1K--10K) videos accounting for over 83\% of instances; growth curves further diverge across tiers, as Viral videos (1M+) exhibit steep early surges with high variance while lower tiers plateau quickly.
The web source distribution in Figure~\ref{fig:stats_dataset} (c) confirms that the three evidence-card dimensions draw from structurally different parts of the web: D1 (topic \& entity) relies on news and official sources, D2 (public discourse) is dominated by forums and social media (over 90\%), and D3 (related content) draws primarily from video platforms and social media.

\section{\textsc{Shorts-Cast}}
\label{sec:method}
\begin{figure}
    \centering
    \includegraphics[width=1\linewidth]{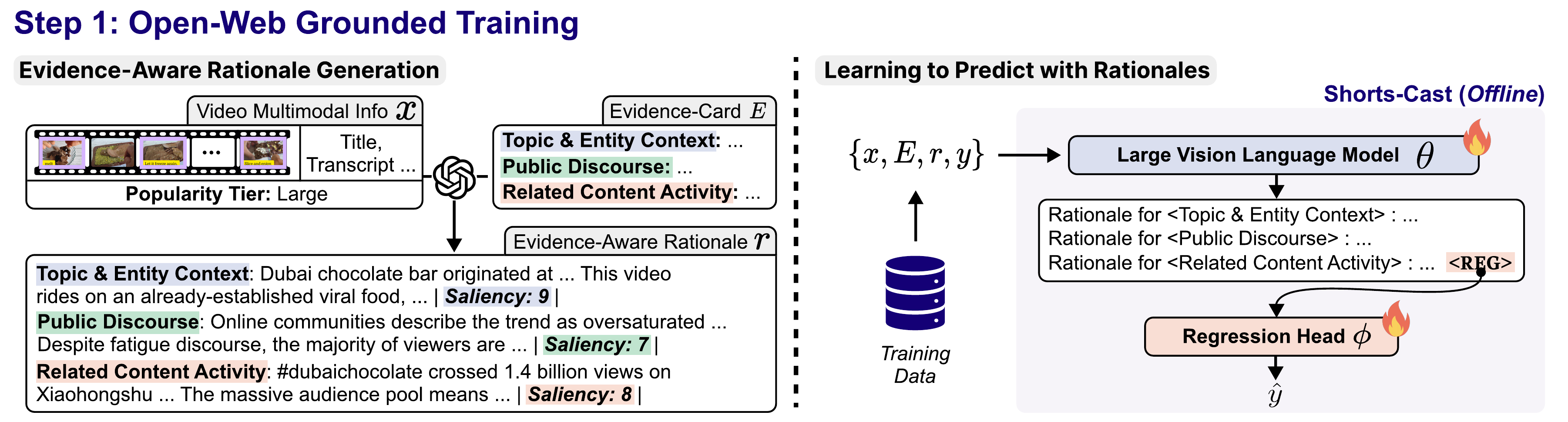}
    \caption{Overview of \proposed Step 1, open-web grounded training.}
    \label{fig:offline_fig}
\end{figure}

\begin{figure}
    \centering
    \includegraphics[width=1\linewidth]{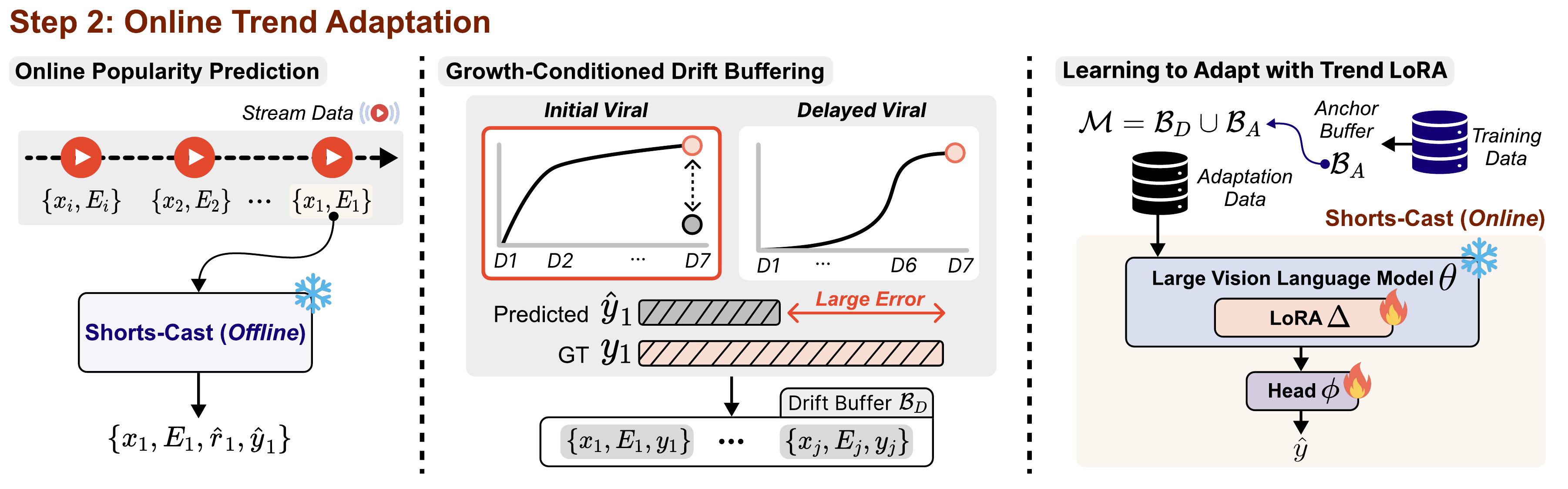}
    \caption{Overview of \proposed Step 2, online trend adaptation.}
    \label{fig:online_fig}
    \vspace{-0.005cm}
\end{figure}

In this section, we present \proposed, a framework that forecasts popularity through open-web evidence-grounded reasoning and adapts to evolving trends at deployment time.
We train \proposed\ on \dataset\ in two steps: (1) training a predictor that first generates an evidence-grounded rationale from the evidence-card, then regresses scalar popularity from the reasoning-conditioned hidden state (Figure~\ref{fig:offline_fig}); and (2)~trend-aware online adaptation that updates the evidence-to-popularity mapping only when delayed-label feedback reveals genuine trend shifts (Figure~\ref{fig:online_fig}).


\subsection{Step 1: Open-Web Grounded Training}
\paragraph{Evidence-aware rationale generation}
Although the evidence-card provides rich contextual signals, directly mapping it to a scalar score does not supervise the intermediate evidence interpretation; the predictor receives no guidance on which evidence dimension should be emphasized or how each dimension contributes to the predicted outcome.
To address this, we train the predictor to generate an evidence-aware rationale before producing the scalar prediction.
Specifically, for each micro-video $V_i$ with multi-modal content $X_i$ (title, transcript, thumbnail, and sampled key frames) and evidence-card $E_i$, we first map the ground-truth raw view count $v_i(T)$ to a coarse popularity tier $c_i = \texttt{tier}(v_i(T)) \in \{\textsc{Micro}, \textsc{Small}, \textsc{Medium}, \textsc{Large}, \textsc{Viral}\}$ with thresholds at 1K, 10K, 100K, and 1M views.
We then prompt GPT-4o-mini~\cite{openai2023gpt4} with $(X_i, E_i, c_i)$ to generate a rationale $r_i = \texttt{LLM}(X_i, E_i, c_i)$ that provides a per-dimension natural-language explanation linking each evidence dimension to the target video's popularity tier.
The rationale additionally assigns each dimension a saliency score (1--10) quantifying how strongly that dimension's web signals contribute to the video's predicted popularity, giving the predictor an explicit signal for weighing which evidence dimensions matter more for a given video.


\paragraph{Learning to predict with rationales}
We define a unified predictor $g_{\theta,\phi}(I_i) = (\hat{r}_i, \hat{y}_i)$, where $I_i = (X_i, E_i)$ denotes the combined input, with a large vision-language model (LVLM) backbone $\theta$ and a lightweight MLP regression head $\phi$.
Given $I_i$, the backbone $\theta$ first generates the rationale $\hat{r}_i$ autoregressively under $p_\theta(r_l \mid r_{<l}, I_i)$.
We then append a special \texttt{<REG>} token at the end of the rationale and feed its hidden state $h_i$ into $\phi$ to produce the scalar prediction $\hat{y}_i = \phi(h_i)$.
Since \texttt{<REG>} appears after the rationale rather than after the raw input, $h_i$ encodes the reasoning the model has just articulated, including which evidence dimensions it deemed salient and why.
Motivated by recent work that couples rationale generation with scalar evaluation~\cite{ankner2024critique, zhang2025mmrlhf, lee2025personalized, yang2026jointreward, zhou2025llavareward}, we jointly optimize rationale quality and prediction accuracy by fine-tuning $\{\theta, \phi\}$ with
\begin{equation}
\mathcal{L}_{\text{Offline}} = \frac{1}{N}\sum_{i=1}^{N}\left[-\frac{1}{|r_i|}\sum_l \log p_\theta(r_l \mid r_{<l}, I_i) + \alpha \cdot (\hat{y}_i - y_i)^2\right],
\end{equation}
where the first term supervises rationale generation, the second term supervises scalar prediction, and $\alpha$ controls their relative weight.
Since both terms share the backbone $\theta$, prediction errors not only correct the regression head $\phi$ but also back-propagate into how the model reasons about evidence, shaping its dimension-level explanations and saliency assignments.
This rationale therefore serves as a readable trace of which evidence dimensions influenced the predicted popularity and how strongly.
We denote the resulting parameters $\{\theta^{(1)}, \phi^{(1)}\}$, which serve as the starting point for deployment.

\subsection{Step 2: Online Trend Adaptation}
\paragraph{Online popularity prediction}
The offline mapping $\{\theta^{(1)}, \phi^{(1)}\}$ is fixed at training time, yet the evidence-card is always fresh at prediction time. 
As trends evolve, the model grows increasingly insensitive to how fresh web evidence relates to current popularity dynamics, interpreting new signals through an outdated mapping.
To adapt this mapping under realistic deployment constraints, we operate on a chronologically ordered video stream with delayed-label feedback. 
Each video $V_i$ is predicted at upload time using its evidence-card $E_i$ collected at that moment, and the ground-truth label $y_i$ becomes visible only after the 7-day prediction horizon elapses.

\paragraph{Growth-Conditioned Drift buffering}
Not all large prediction errors indicate that the learned mapping is outdated.
Since video view counts follow a long-tailed distribution, large errors frequently arise from regression noise on extreme samples rather than from genuine trend shifts.
A reliable drift filter must therefore separate \textit{why} the predictor failed, not merely \textit{how much}.
Prior work~\cite{cao2021cascade} shows that content-based predictors fail most severely on videos with abnormal growth curves, where popularity is shaped by external trend dynamics rather than intrinsic content quality.
We therefore introduce \textit{Growth-Conditioned Drift}, a filtering strategy that combines the view-growth pattern with error magnitude as a joint drift signal, ensuring that only trend-driven failures trigger adaptation.

Concretely, when the label $y_i$ arrives, we compute the scalar error $e_i = \hat{y}_i - y_i$ and the growth ratio $\gamma_i = v_i(2) / v_i(7)$, the fraction of 7-day cumulative views already accumulated by day~2.
Following the growth-curve taxonomy of~\cite{france2016characterizing}, we calibrate percentile thresholds $\gamma_{\text{low}}$ and $\gamma_{\text{high}}$ on $\mathcal{D}_{\text{val}}$ so that $\gamma_i > \gamma_{\text{high}}$ captures \textit{initial viral} growth (rapid early surge then plateau) and $\gamma_i < \gamma_{\text{low}}$ captures \textit{delayed viral} growth (low initial views followed by a late burst)~\cite{rizoiu2017hip}.
Videos within the normal range ($\gamma_i \in (\gamma_{\text{low}}, \gamma_{\text{high}})$) are excluded regardless of error magnitude.
We define the growth gate $G_i = \mathbb{I}[\gamma_i \notin (\gamma_{\text{low}}, \gamma_{\text{high}})]$ and the drift indicator $d_i = G_i \cdot \mathbb{I}[|e_i| > \delta_y]$, where $\delta_y$ is an error threshold calibrated on $\mathcal{D}_{\text{val}}$.
Samples with $d_i=1$ are inserted into the drift buffer, $\mathcal{B}_D \leftarrow \mathcal{B}_D \cup \{(I_i, y_i)\}$.

\paragraph{Learning to adapt with trend LoRA}
The adaptation objective is to correct the drifted evidence-to-popularity mapping while preserving the reasoning ability learned in Step 1.
We freeze the offline backbone $\theta^{(1)}$ and attach lightweight LoRA adapters~\cite{hu2022lora} to all attention projections, yielding the trainable parameter set $\{\Delta, \phi\}$ in Step 2; the LM head remains frozen, preserving the saliency calibration learned in Step 1.
Adaptation is triggered when the drift buffer $\mathcal{B}_D$ accumulates $M$ new samples since the last update.
Upon triggering, we construct the adaptation set $\mathcal{M} = \mathcal{B}_D \cup \mathcal{B}_A$, where $\mathcal{B}_A$ is an anchor buffer randomly sampled from the training set to prevent catastrophic forgetting, and update $\{\Delta, \phi\}$ with scalar regression loss only:
\begin{equation}
\mathcal{L}_{\text{Online}} = \frac{1}{|\mathcal{M}|}\sum_{(I,y) \in \mathcal{M}} (\hat{y} - y)^2, \quad \hat{y} = g_{\theta^{(1)}+\Delta,\,\phi}(I)
\end{equation}
We exclude the next-token prediction (NTP) loss from online adaptation, as the frozen LM head already preserves the rationale generation ability learned in Step 1.
After each update, the new-sample counter is reset and the adapted predictor is applied to all subsequent incoming videos, progressively aligning the evidence-to-popularity mapping with the current trends.

\section{Experiments}
\label{sec:exp}

\begin{wraptable}{r}{0.5\columnwidth}
\vspace{-1.3cm}
\caption{Offline evaluation results on \dataset. \texttt{<REG>} denotes the regression head.}
\label{tab:offline_main}
\centering
\renewcommand{\arraystretch}{1.05}
\setlength{\tabcolsep}{1.5pt}
\resizebox{0.5\columnwidth}{!}{
\begin{tabular}{@{}lcccc@{}}
\toprule
\textbf{Method} & nMSE ($\downarrow$) & MAE ($\downarrow$) & SRC  ($\uparrow$) & PCC ($\uparrow$) \\
\midrule
\multicolumn{5}{@{}l}{\cellcolor{gray!12}\textit{\textbf{Video Content-Only}}} \\
w/o \texttt{<REG>} & 1.049 & 2.933 & 0.371 & 0.379 \\
w/ \texttt{<REG>} & 0.894 & 2.689 & 0.448 & 0.396 \\
\midrule
\multicolumn{5}{@{}l}{\cellcolor{gray!12}\textit{\textbf{Video Corpus Retrieval-Augmented}}} \\
w/o \texttt{<REG>} & 0.899 & 2.709 & 0.409 & 0.472 \\
w/ \texttt{<REG>} & 0.878 & 2.563 & 0.425 & 0.398 \\
\midrule
\multicolumn{5}{@{}l}{\cellcolor{gray!12}\textit{\textbf{Open-Web Grounded}}} \\
Raw Web Snippet w/o \texttt{<REG>} & 1.025 & 3.306 & 0.336 & 0.221 \\
Raw Web Snippet w/ \texttt{<REG>} & 0.923 & 2.791 & 0.228 & 0.347 \\
Evidence-card w/o \texttt{<REG>} & 0.905 & 2.613 & 0.368 & 0.407 \\
Evidence-card w/ \texttt{<REG>} & \underline{0.826} & \underline{2.542} & \underline{0.518} & \underline{0.491} \\
\textbf{\proposed (\textit{Offline})} & \textbf{0.701} & \textbf{2.453} & \textbf{0.612} & \textbf{0.594} \\
\bottomrule
\end{tabular}%
}
\vspace{-0.97cm}
\end{wraptable}

\subsection{Experimental setup}
\paragraph{Dataset}
All experiments are conducted on \dataset. We split chronologically by upload time into train/val/stream/test at 7:1:1:1. $\mathcal{D}_{\text{stream}}$ is reserved for the prequential online protocol (Section \S~\ref{subsec:online}).

\paragraph{Metrics}
Following prior works~\cite{ye2025mvp, zhong2024mmra, cheng2025seeing, cheng2024ragtrans, chen2025echoes}, we evaluate with four metrics: normalized mean squared error (nMSE), mean absolute error (MAE), Spearman's rank correlation (SRC), and Pearson correlation coefficient (PCC).
nMSE normalizes MSE by target variance, providing a scale-invariant error measure that remains comparable across different popularity distributions.

\paragraph{Implementation details}

All methods use Qwen3-VL-8B-Instruct~\cite{yang2025qwen3}. Step 1 fine-tunes $\{\theta,\phi\}$ with AdamW using joint NTP and MSE losses; Step 2 freezes $\theta^{(1)}$ and trains attention-projection LoRA adapters.
Thresholds and buffers are calibrated on $\mathcal{D}_{\text{val}}$; hyperparameters are in Appendix~\ref{sec:apn_b}.

\subsection{Offline Evaluation}
\label{subsec:offline}
\paragraph{Baselines}
We compare three input settings, each evaluated with and without the \texttt{<REG>} regression head:
\textbf{(1) Content-Only}, using only video multi-modal content $X_i$; \textbf{(2) Video Corpus Retrieval-Augmented}, adding top-5 historical neighbors from $\mathcal{D}_{\text{train}}$~\cite{zhong2024mmra, chen2025echoes, cheng2025incontext}; and \textbf{(3) Web-Grounded}, using either raw top-5 search snippets or our structured evidence card. 
The evidence-Card w/~\texttt{<REG>} variant shares \proposed's input but predicts without generating rationales.


\begin{wraptable}{r}{0.45\columnwidth}
\vspace{-0.4cm} 
\caption{Effect of evidence-card}
\centering
\resizebox{0.45\columnwidth}{!}{
\begin{tabular}{lcccc}
\toprule
\textbf{Method} & \textbf{nMSE} ($\downarrow$) & \textbf{MAE} ($\downarrow$) & \textbf{SRC} ($\uparrow$) & \textbf{PCC} ($\uparrow$) \\
\midrule
MMRA~\cite{zhong2024mmra} & 0.807 & 2.915 & 0.480 & 0.458 \\
\rowcolor{black!7}\textbf{w/ EC} & \textbf{0.786} & \textbf{2.840} & \textbf{0.494} & \textbf{0.470} \\
\midrule
EvoPro~\cite{chen2025echoes} & 0.822 & 2.894 & 0.458 & 0.425 \\
\rowcolor{black!7}\textbf{w/ EC} & \textbf{0.816} & \textbf{2.872} & \textbf{0.466} & \textbf{0.432} \\
\midrule
ICPF~\cite{cheng2025incontext} & 0.794 & 2.817 & 0.478 & 0.454 \\
\rowcolor{black!7}\textbf{w/ EC} & \textbf{0.787} & \textbf{2.807} & \textbf{0.489} & \textbf{0.462} \\
\bottomrule
\end{tabular}
}
\label{tab:prior_methods}
\vspace{-0.4cm}
\end{wraptable}


\paragraph{Results}
As shown in Table~\ref{tab:offline_main}, the structured evidence-card outperforms video-corpus retrieval under the same \texttt{<REG>} head, confirming that temporally aligned open-web context provides stronger predictive signals than historical platform-internal neighbors.
This advantage, however, depends on structure: raw web snippets degrade performance below the content-only baseline, showing that unstructured web text adds noise rather than reliable context.
This structured advantage also generalizes to existing methods: Table~\ref{tab:prior_methods} shows that augmenting prior MVPP methods with our evidence-card consistently improves their performance,
indicating that structured web context serves as a broadly useful external signal independent of the prediction framework.
The stronger performance of \proposed~(\textit{Offline}) then suggests that these signals are most beneficial when they are not merely appended to the input, but converted into a reasoning-conditioned representation.
Overall, the offline results demonstrate that open-web grounding requires both structured evidence construction and evidence-aware reasoning, rather than simple context augmentation. 

\subsection{Online Evaluation}
\label{subsec:online}
\paragraph{Setting}
All online methods are initialized from the \proposed~(\textit{Offline}) checkpoint and consume $\mathcal{D}_{\text{stream}}$ in chronological order: each video is predicted at upload time, and its day-7 label becomes visible only after the 7-day horizon elapses, while $\mathcal{D}_{\text{train}}$ serves as the anchor-buffer source.
We report two metrics: \textit{Prequential} scores each prediction over $\mathcal{D}_{\text{stream}}$ before any adaptation from that sample, and \textit{Test} evaluates the final adapted model on $\mathcal{D}_{\text{test}}$ after the stream concludes.

\begin{table*}[!t]
\caption{Online evaluation on \dataset. Prequential scores each video before adaptation; Test evaluates the final model after the stream.}
\centering
\scriptsize
\renewcommand{\arraystretch}{1.02}
\setlength{\tabcolsep}{2pt}
\resizebox{0.99\textwidth}{!}{
\begin{tabular}{lcccccccc}
\toprule
 & \multicolumn{4}{c}{\textbf{Prequential}} & \multicolumn{4}{c}{\textbf{Test}} \\
\cmidrule(lr){2-5} \cmidrule(lr){6-9}
\multirow{-2.5}{*}{\textbf{Method}}  
& nMSE ($\downarrow$) & MAE ($\downarrow$) & SRC ($\uparrow$) & PCC ($\uparrow$) 
& nMSE ($\downarrow$) & MAE ($\downarrow$) & SRC ($\uparrow$) & PCC ($\uparrow$) \\
\midrule
\multicolumn{9}{@{}l}{\cellcolor{gray!12}\textit{\textbf{No Adaptation}}} \\
\proposed (\textit{Offline}) & - & - & - & - & 0.701 & \underline{2.453} & 0.612 & 0.594 \\
\midrule
\multicolumn{9}{@{}l}{\cellcolor{gray!12}\textit{\textbf{Non-Parametric Adaptation}}} \\
Cache-kNN                      & \underline{0.932} & \underline{2.995} & \underline{0.554} & \underline{0.512} & 0.885 & 2.751 & 0.591 & 0.546 \\
Temporal-Decay Residual Memory & 0.958 & 3.042 & 0.551 & 0.507 & 0.667 & 2.498 & 0.592 & 0.582 \\
\midrule
\multicolumn{9}{@{}l}{\cellcolor{gray!12}\textit{\textbf{Parametric Adaptation}}} \\
OGD-LoRA                     & 0.960 & 3.043 & 0.547 & 0.504 & 0.673 & 2.685 & \underline{0.642} & 0.623 \\
ER-LoRA                      & 0.952 & 3.021 & 0.551 & 0.508 & \underline{0.624} & 2.498 & 0.637 & \underline{0.624} \\
\textbf{\proposed (\textit{Online})} & \textbf{0.730} & \textbf{2.678} & \textbf{0.583} & \textbf{0.563} & \textbf{0.592} & \textbf{2.258} &  \textbf{0.677} & \textbf{0.665} \\
\bottomrule
\end{tabular}
}
\label{tab:online_main}
\vspace{-0.1cm}
\end{table*}

\paragraph{Baselines}
All baselines share the \proposed~(\textit{Offline}) initialization and differ in how they react to delayed labels.
\textbf{(1) Non-parametric} methods adjust outputs without weight updates:
Cache-$k$NN~\cite{khandelwalgeneralization, bhardwaj-etal-2023-adaptation}
retrieves revealed neighbors, and Temporal-Decay Residual Memory~\cite{zhang2023adanpc, lyu2026ts} corrects predictions with time-decayed per-topic residuals.
\textbf{(2) Parametric} methods update LoRA adapters on every revealed batch: OGD-LoRA~\cite{wei2025onlinelora, wang2023orthogonal} without filtering, and ER-LoRA~\cite{zhang2022a} with uniform experience replay.
\proposed instead triggers updates only on growth-conditioned drift samples. 


\paragraph{Results}
Table~\ref{tab:online_main} shows that non-parametric methods adjust outputs without changing the context-to-popularity mapping, limiting their capacity to correct systematic failures (Cache-$k$NN even degrades test nMSE from 0.701 to 0.885).
Parametric methods update the same LoRA architecture as \proposed and improve test nMSE (0.673, 0.624), yet treating every delayed label equally destabilizes the mapping mid-stream, producing the weakest prequential scores (0.960, 0.952).
By contrast, \proposed~triggers adaptation only when a large prediction error co-occurs with an abnormal growth pattern, selectively retaining the trend-driven failures that signal a genuine shift in the trend landscape.
This trend-aware filtering yields the strongest prequential nMSE (0.730) while achieving the best test performance (nMSE 0.592, SRC 0.677), confirming that conditioning adaptation on trend-typed
failures both stabilizes the mapping mid-stream and generalizes to future videos.

\section{Discussion}
\label{sec:discuss}
\begin{figure}[t]
    \centering
    \includegraphics[width=0.99\linewidth]{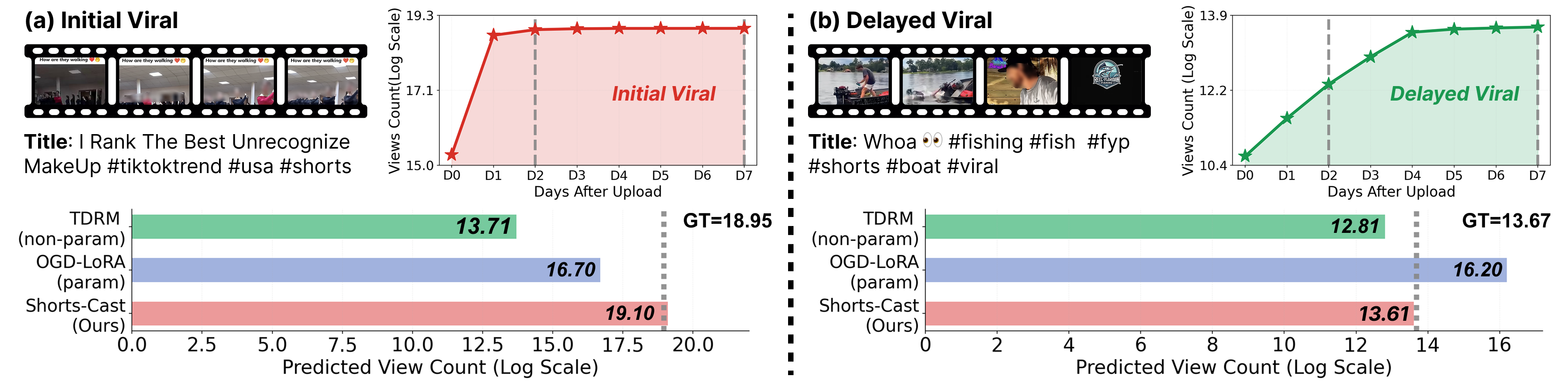}
    \caption{Case study of \proposed on initial viral \textbf{(Left)} and delayed viral \textbf{(Right)} videos, showing predicted view counts against the ground truth across online baselines.}
    \label{fig:case_study}
    \vspace{-0.4cm}
\end{figure}

\begin{wrapfigure}{r}{0.35\columnwidth}
\centering
\vspace{-13pt}
\captionof{table}{Performance across evidence-card snapshots}
\label{tab:ec_refresh}
\vspace{-4pt}
\resizebox{0.35\columnwidth}{!}{%
\begin{tabular}{l cccc}
\toprule
$t$ & nMSE $\downarrow$ & MAE $\downarrow$ & SRC $\uparrow$ & PCC $\uparrow$ \\
\midrule
0 & 0.851 & 2.696 & 0.610 & 0.563 \\
1 & \underline{0.729} & \underline{2.525} & \underline{0.697} & \underline{0.671} \\
2 & \textbf{0.696} & \textbf{2.469} & \textbf{0.721} & \textbf{0.694} \\
\bottomrule
\end{tabular}%
}
\par\vspace{9pt}           
\includegraphics[width=0.35\columnwidth]{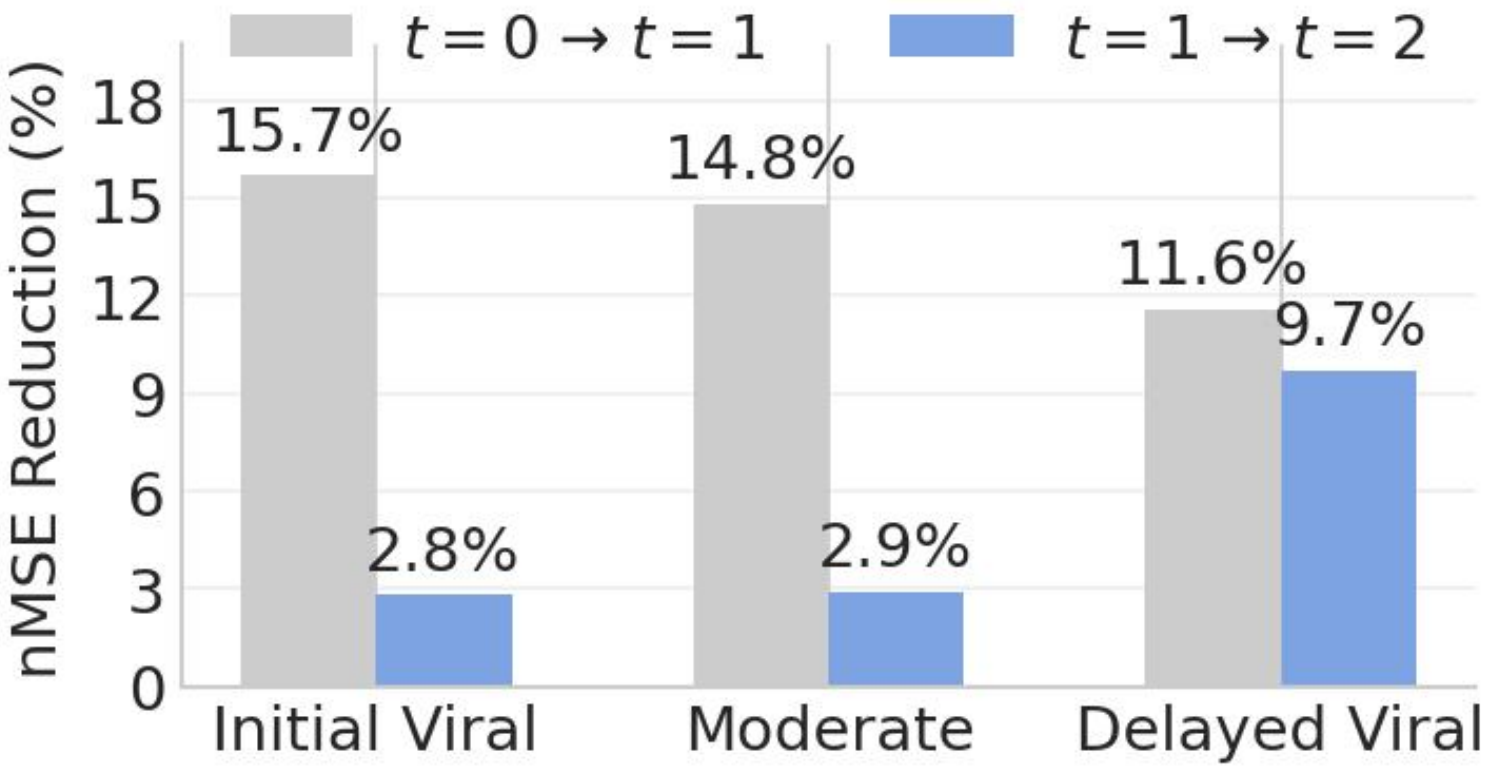}
\vspace{-7pt}               
\captionof{figure}{nMSE across evidence-card snapshots by growth type.}
\label{fig:ec_analysis}
\vspace{-18pt}
\end{wrapfigure}

\paragraph{Effect of evidence-card refresh}
We fix \proposed and vary only the observation snapshot $t \in \{0, 1, 2\}$ at which the evidence-card is collected.
Table~\ref{tab:ec_refresh} shows that later evidence consistently improves all metrics, with nMSE dropping from 0.851 ($t$=0) to 0.696 ($t$=2).
Figure~\ref{fig:ec_analysis} further reveals that this gain distributes unevenly across growth types: for \textit{initial viral} and \textit{typical} videos, nearly all nMSE reduction occurs from $t$=0 to $t$=1 ($<$3\% additional gain afterward), whereas \textit{delayed viral} videos retain a 9.7\% second-step reduction, nearly matching their first step.
This suggests that evidence-cards are most valuable when their collection timing aligns with the video's growth dynamics: videos that peak early expose predictive context quickly, whereas late-growing videos depend on discourse that materializes only after upload.

\textbf{Case study}
Figure~\ref{fig:case_study} reveals that all online baselines under-predict the initial viral video by 2 to 5 log-scale points, despite their respective adaptation mechanisms. 
The failure is more acute in the delayed viral case, where OGD-LoRA over-predicts by 2.5 points; updating on every delayed label without distinguishing growth type causes the mapping to overcompensate in the opposite direction. 
\proposed most closely approaches the ground truth in both cases, predicting within 0.15 and 0.06 log-scale points respectively. 
This highlights that growth-conditioned drift filtering, which admits samples into the drift buffer by the growth ratio $\gamma_i$ rather than error magnitude alone, resolves the inter-growth-type trade-off that indiscriminate adaptation creates.



\section{Conclusion}
\label{sec:conclusion}

In this work, we reformulate micro-video popularity prediction as open-web grounded prediction, introducing \dataset and \proposed to leverage structured external web context and adapt to shifting trends at deployment.
Our experiments demonstrate that open-web context is essential for accurate prediction, yet its value depends critically on how it is delivered.
Furthermore, reacting to every prediction error leads to overcorrection; conditioning updates on view-growth patterns instead enables the model to distinguish genuine trend shifts from routine long-tail noise.
We believe this work establishes a foundation for leveraging the open web as a structured, temporally aligned signal source for content popularity forecasting.

{
\small
\bibliographystyle{unsrtnat}  
\bibliography{reference}
}


\newpage
\appendix
\section{Limitations and Broader Impacts}
\label{sec:apn_limitations}
\paragraph{Limitations}
While our results confirm the value of open-web grounding and trend-aware adaptation for MVPP, we acknowledge the following limitations of the current work.

First, \dataset is collected exclusively from YouTube Shorts and restricted to English-language videos. Short-form platforms differ substantially in their recommendation algorithms, user behavior, and content exposure mechanisms, so the evidence-to-popularity mapping learned on YouTube Shorts may not transfer directly to platforms such as TikTok, Kuaishou, or Instagram Reels. 
Extending the collection pipeline to multiple platforms and languages would enable cross-platform evaluation and reveal how platform-specific exposure biases interact with open-web context.

Second, although Growth-Conditioned Drift filtering provides a principled criterion for separating trend-driven failures from regression noise, the underlying growth-curve taxonomy~\cite{france2016characterizing} categorizes videos into a fixed set of growth types calibrated on the validation set. 
In practice, user-generated short-form content exhibits far more diverse and rapidly shifting trend dynamics than a static taxonomy can capture; new growth patterns that fall outside the calibrated thresholds may go undetected, and the optimal moment to trigger adaptation can vary substantially across trend regimes. 
Exploring more adaptive drift detection mechanisms, such as learnable gating functions or non-stationary threshold calibration, is a necessary direction for improving the timeliness and precision of online adaptation.

Lastly, we adopt a fixed 7-day prediction horizon based on prior work~\cite{xu2025smtpd}, which shows that cumulative view counts of most micro-videos converge within roughly one week after upload. However, videos responding to slow-building trends or seasonal events can continue accumulating views well beyond this window, and the 7-day delayed-label requirement also means that online adaptation is structurally one week behind the fastest trend cycles.
Collecting popularity growth curves over longer horizons (e.g., 14, 21, or 30 days) would support multi-horizon prediction and reduce adaptation lag by allowing shorter-horizon labels to trigger earlier updates.

\paragraph{Broader impacts}
\dataset and \proposed can support research on web-grounded multimodal reasoning, content recommendation, and platform analytics by making the contextual signals behind early popularity more transparent and analyzable. 
At the same time, popularity prediction technologies carry potential negative impacts: they may be exploited for engagement farming, amplification of misleading or low-quality content, or to gain unfair advantages over creators without similar tooling, and any biases in topic, language, or platform coverage of \dataset can propagate to downstream models. 
To mitigate these risks, we restrict \dataset to research use, redistribute only video identifiers and derived metadata rather than raw video content (in line with YouTube's Terms of Service).


\section{\dataset Details}
\label{sec:apn_a}

\subsection{Video Collection}
\label{subsec:appendix_video_collection}
\begin{table}[t]
\centering
\small
\setlength{\tabcolsep}{4pt}
\renewcommand{\arraystretch}{1.15}
\caption{Seed query construction: 10 trend features $\times$ 17 topics $\times$ 10 sub-topics, yielding 1{,}700 templates of the form \textit{``\{Trend Feature\} \{Topic\} \{Sub-Topic\} \#shorts''}.}
\label{tab:seed_queries}
\vspace{4pt}
\begin{tabular}{@{}p{0.14\linewidth} p{0.17\linewidth} p{0.61\linewidth}@{}}
\toprule
\textbf{Trend feature} & \textbf{Topic} & \textbf{Sub-topics} \\
\midrule
Hot        & Animals        & Dog, Wildlife, Cat, Fish, Birds, Insect, Snakes, Pet, Amphibians, Reptile \\
Trending   & Education      & School, Club, Teacher, Speaking, Listening, Writing, Presentation, Math, Computer, Teamwork \\
Viral      & Health         & Mental, Injury, Medication, Digestive health, Dental, Optical, Reproductive, Skin, Brain health, Cardiac \\
Latest     & Travel         & Museum, Park, Sea, Beach, Mountain, Lake, Hotel, Resort, Camping, Hiking \\
Best       & Movies         & Action movie, Comedy, Romance, Science fiction, Horror, Drama, Cartoon, Documentary, Adventure, Crime \\
Highlights & Cooking        & Broiling, Grilling, Roasting, Baking, Sauteing, Boiling, Steaming, Poaching, Simmering, Stewing \\
Recent     & Job            & Manager, Researcher, Chef, Police, Lawyer, Salesman, Mechanic, Banker, Doctor, Waiter \\
Top        & Electronics    & Laptop, TV, Phone, Software, Internet, Camera, Audio, Headphone, Hardware, Monitor \\
New        & Art            & Crafts, Photography, Painting, Collection, Drawing, Digital art, Sculpting, Pottery, Glass craft, Calligraphy \\
Popular    & Personal Style & Grooming, Fashion, Personal Hygiene, Tattoos, Scarf, Hair Style, Makeup, Dressing, Tie, Formal \\
           & Clothes        & Sweater, Jeans, Shirt, Socks, Coat, Pants, Hat, Gloves, Dress, Shoes \\
           & Sports         & Outdoor recreation, Team sports, Tennis, Football, Basketball, Climbing, Skiing, Swimming, Fishing, Yoga \\
           & House          & Building, Garden, Pool, Bathroom, Bedroom, Kitchen, Repairment, Moving, Decoration, Furniture \\
           & Food           & Fruit, Vegetable, Drink, Meat, Seafood, Snacks, Dessert, Breakfast, Lunch, Dinner \\
           & Holiday        & Halloween, Christmas, Labor day, Thanksgiving, Valentine's day, Mother's day, Birthday, National day, New year, Father's day \\
           & Transportation & Car, Train, Bus, Boat, Bike, Airplane, Motorcycle, Truck, Trailer, Scooter \\
           & Hobbies        & Dancing, Singing, Playing cards, Reading, Chess, Board games, Team games, Volunteer work, Instrument, Exercise \\
\bottomrule
\end{tabular}
\end{table}

\paragraph{Candidate retrieval}
All candidate videos and their metadata are collected through the YouTube Data API v3, accessed via a Google Cloud project\footnote{\url{https://console.cloud.google.com/}}. 
We issue queries with each of the 1,700 seed templates listed in Table~\ref{tab:seed_queries}, restricting results to short-form videos and retaining only the fields necessary for downstream processing (video ID, title, channel ID, upload timestamp, and view-count). 

\paragraph{Candidate filtering}
In addition to the metadata-completeness and English-language filters described in Section~\ref{sec:dataset}, we apply a per-channel cap that restricts the number of videos retained from any single channel. 
This prevents the dataset from being dominated by a small number of large or prolific channels and ensures diverse video sources across topics and creators.

\paragraph{Popularity tracking}
Each surviving candidate is registered into a polling queue, and we query the YouTube Data API for its cumulative view count at every 24-hour boundary relative to the upload timestamp until Day~7. 
Videos that become deleted, privated, or otherwise inaccessible during the tracking window are flagged and excluded from the final dataset to avoid truncated popularity growth curves.

\subsection{Evidence-Card Analysis}
\label{subsec:appendix_ec_construction}
\paragraph{Evidence-card construction}
We instantiate the search-augmented generation procedure described in Section~\ref{sec:dataset} with grok-4.1-fast-reasoning, providing each video's multi-modal content $X_i$ and the observation day $t$ as input, and constraining the search scope by excluding the \texttt{youtube.com} domain to prevent platform leakage.
Each evidence-card is returned as a JSON object with the three dimensions D1--D3 and accompanying source indices.
The full prompt template is shown in Table~\ref{tab:evidence_card_prompt}.

\begin{table*}
\small
\caption{Prompt for evidence-card generation.}
\label{tab:evidence_card_prompt}
\centering
\setlength{\tabcolsep}{4pt}
\renewcommand{\arraystretch}{1.05}
\begin{tabular}{p{14cm}}
\toprule
\textbf{Evidence-card generation prompt} \\
\midrule
\textcolor{teal}{\textbf{[Task]}}\\
You are an open-web analyst for micro-video popularity prediction. Given a target YouTube Shorts video and an observation day, search the open web and produce a structured Evidence Card that summarizes external context across three dimensions: \textbf{D1 Topic \& Entity Context} (factual background, events, entities, timelines), \textbf{D2 Public Discourse} (reactions, sentiment, controversies, fandom, community discussion), and \textbf{D3 Related Content Activity} (similar or competing short-form content that may amplify or dilute attention to the target video). \\[2pt]
\textcolor{teal}{\textbf{[Inputs \& Constraints]}}\\
Use only the target video's multimodal content (title; four key frames at zero, twenty-five, fifty, and seventy-five percent positions, where the zero percent frame is the thumbnail; and the ASR transcript split into segments temporally aligned with each frame) together with the observation day. Do not use or request channel identity, subscriber count or tier, source query, source region, country, platform-internal recommendation signals, the video's future views, likes, or comments, the raw day-seven view count, or the coarse popularity tier (the tier is consumed only by evidence-aware rationale generation). \textbf{Temporal integrity:} restrict open-web sources to those available by the observation day when source dates are present; if a source date is missing, keep the source only when useful for context and never infer a future date. \textbf{Platform leakage guard:} exclude \texttt{youtube.com}, \texttt{youtu.be}, and \texttt{m.youtube.com} via \texttt{excluded\_domains} so platform-internal popularity signals cannot enter the card. \\[2pt]
\textcolor{teal}{\textbf{[User Prompt Template]}}\\
Title: \texttt{\{title\}}\quad Observation day: \texttt{\{observation\_day\}}\\
Interleaved frames and ASR (in temporal order):\\
\quad[Frame 0\% / thumbnail]: \texttt{\{frame\_0\}}\quad ASR (0--25\%): \texttt{\{asr\_seg\_0\}}\\
\quad[Frame 25\%]: \texttt{\{frame\_25\}}\quad ASR (25--50\%): \texttt{\{asr\_seg\_25\}}\\
\quad[Frame 50\%]: \texttt{\{frame\_50\}}\quad ASR (50--75\%): \texttt{\{asr\_seg\_50\}}\\
\quad[Frame 75\%]: \texttt{\{frame\_75\}}\quad ASR (75--100\%): \texttt{\{asr\_seg\_75\}} \\[2pt]
\textcolor{teal}{\textbf{[Output Format \& Schema]}}\\
Return one valid JSON object. Each dimension contains a concise \texttt{evidence}, a list of \texttt{source\_ids}, and a \texttt{source\_index} where each source has \texttt{id}, \texttt{source}, \texttt{type}, \texttt{url}, and \texttt{date} when available. \\
\texttt{\{"evidence\_card": \{"topic\_entity\_context": \{"evidence": <str>, "source\_ids": [<str>], "source\_index": [<source>]\}, "public\_discourse": \{...\}, "related\_content\_activity": \{...\}\}\}} \\
\bottomrule
\end{tabular}
\end{table*}

\paragraph{Evidence-card cost}
Evidence-card construction is performed through the xAI API with grok-4.1-fast-reasoning.
Across the three observation snapshots ($t \in \{0,1,2\}$) collected for each video, the full pipeline issues $153{,}628$ search-augmented requests, consumes approximately $2.60$B tokens, and incurs \$$1{,}217.82$ in total API spend (Table~\ref{tab:ec_cost}).
On a per-card basis this averages to \$$0.028$ ($3.51$ requests, $59.4$K tokens), and at the per-video level (three snapshots per video) to \$$0.083$.

\begin{table}[h]
\caption{Evidence-card construction cost using the xAI API with grok-4.1-fast-reasoning.}
\vspace{0.2cm}
\label{tab:ec_cost}
\centering
\small
\renewcommand{\arraystretch}{1.05}
\setlength{\tabcolsep}{6pt}
\begin{tabular}{lrrr}
\toprule
\textbf{Granularity} & \textbf{Requests} & \textbf{Tokens} & \textbf{Cost (USD)} \\
\midrule
Total                    & $153{,}628$ & $2.60$B   & \$$1{,}217.82$ \\
Per evidence-card        & $3.51$      & $59.4$K   & \$$0.0278$ \\
Per video (3 snapshots)  & $10.53$     & $178$K    & \$$0.0834$ \\
\bottomrule
\end{tabular}
\end{table}

\begin{figure}[h]
\centering
\includegraphics[width=0.99\linewidth]{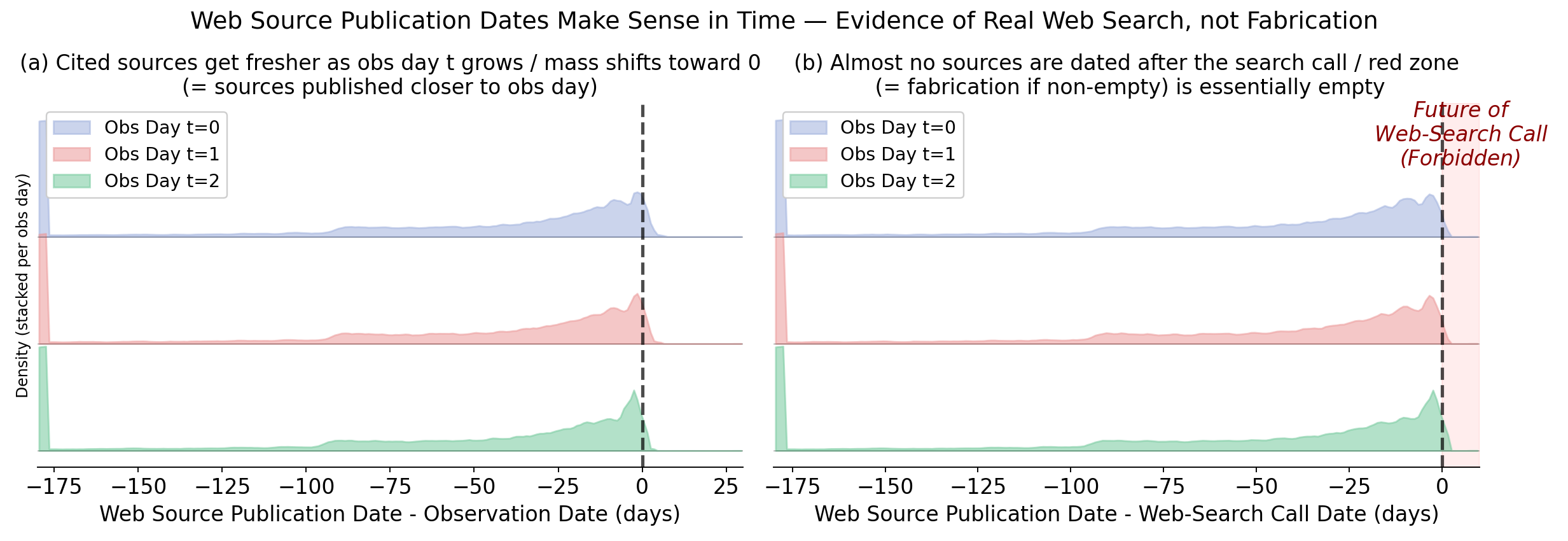}
\caption{Distribution of cited-source publication dates relative to \textbf{(Left)} the snapshot observation time and \textbf{(Right)} the LLM's web-search call time, stacked across observation days $t \in \{0, 1, 2\}$. The red region marks dates after the reference time; in the right panel only $0.03\%$ of sources fall there.}
\label{fig:p1_source_date}
\end{figure}

\paragraph{Source date integrity}
A residual leakage concern beyond YouTube-domain exclusion is that search-augmented LLM may bypass live search and synthesize evidence-card content directly from its pretrained knowledge of viral topics.
We probe this by examining when the cited source pages were published, separately for each observation day.
Figure~\ref{fig:p1_source_date} plots, for each cited source, how many days its publication date precedes ($<$0) or follows ($>$0) two reference times: the snapshot observation time \textbf{(Left)} and the LLM's actual search-call time \textbf{(Right)}, across observation days $t \in \{0, 1, 2\}$.
The \textbf{(Left)} panel exhibits a small rightward mass that simply reflects the unavoidable lag between snapshot time and data collection, so we use the search-call reference in \textbf{(Right)} as the actual leakage test.
The (Right) distribution remains one-sided in every snapshot (only $0.03\%$ of sources, $83$ of $272{,}707$, are dated after the search-call time) and shifts rightward as $t$ grows, with later snapshots citing more recent pages.
The rightward shift is the discriminating signal here: a parametric store frozen at LLM training time cannot selectively introduce pages indexed between $t{=}0$ and $t{=}2$ of a specific video, so this snapshot-conditioned evolution must come from fresh web retrieval rather than parametric recall.

\begin{figure}[h]
\centering
\includegraphics[width=0.99\linewidth]{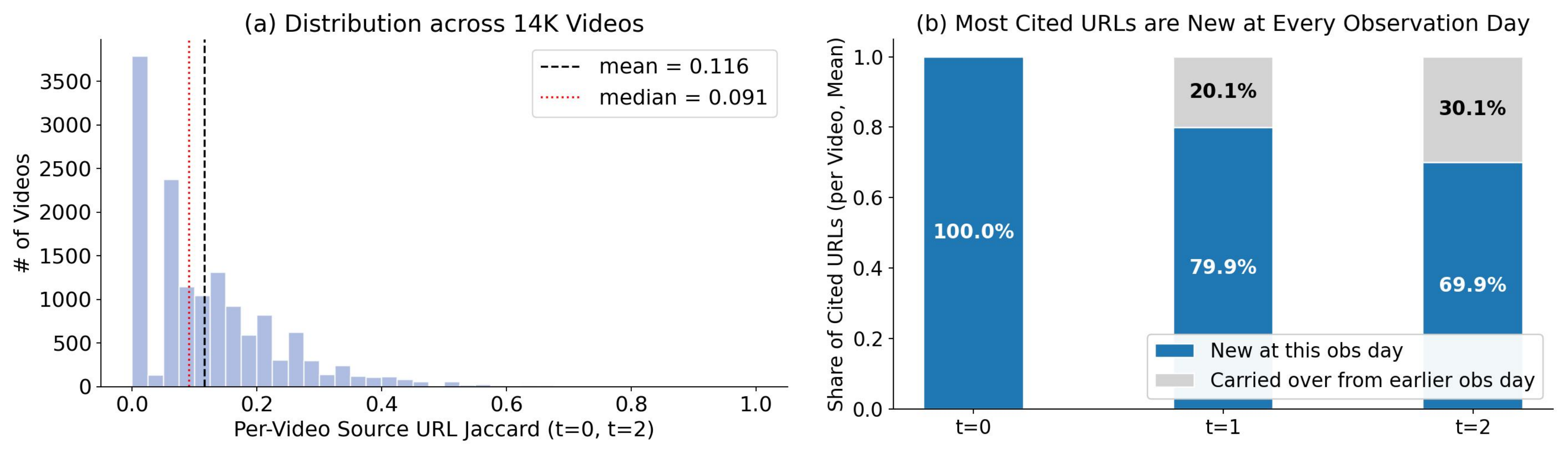}
\caption{Search freshness across observation snapshots. \textbf{Left}: per-video URL-set Jaccard similarity (mean Jaccard heatmap across snapshot pairs and per-video Jaccard$(t{=}0, t{=}2)$ distribution). \textbf{Right}: average share of new (unseen at any earlier snapshot) versus carryover URLs at each observation day.}
\label{fig:search_freshness}
\end{figure}

\paragraph{Search freshness across snapshots}
The date analysis rules out parametric recall at the temporal level, but the LLM could in principle still return the same canonical URL set at every snapshot from memory.
We test this directly by measuring per-video URL-set overlap across snapshots.
Figure~\ref{fig:search_freshness} \textbf{(Left)} reports the mean Jaccard between off-diagonal snapshot pairs at approximately $0.12$, with the per-video Jaccard$(t{=}0, t{=}2)$ distribution concentrated near zero (median $0.091$, mean $0.116$) and roughly $4{,}000$ videos exhibiting fully disjoint URL sets.
Figure~\ref{fig:search_freshness} \textbf{(Right)} further shows that $79.9\%$ of URLs at $t{=}1$ and $69.9\%$ at $t{=}2$ are unseen at the previous snapshot, with carryover rising modestly from $20.1\%$ to $30.1\%$ as canonical pages accumulate.
A static parametric store would yield near-identical URL sets across snapshots for the same input video; the observed near-disjoint URL sets and consistently high new-URL intake therefore confirm that the URLs themselves originate from a fresh search call at each $t$, completing the case against parametric leakage.

\subsection{Channel-size Distribution}
\label{subsec:appendix_stats_details}
\begin{figure}[h]
\centering
\includegraphics[width=0.99\linewidth]{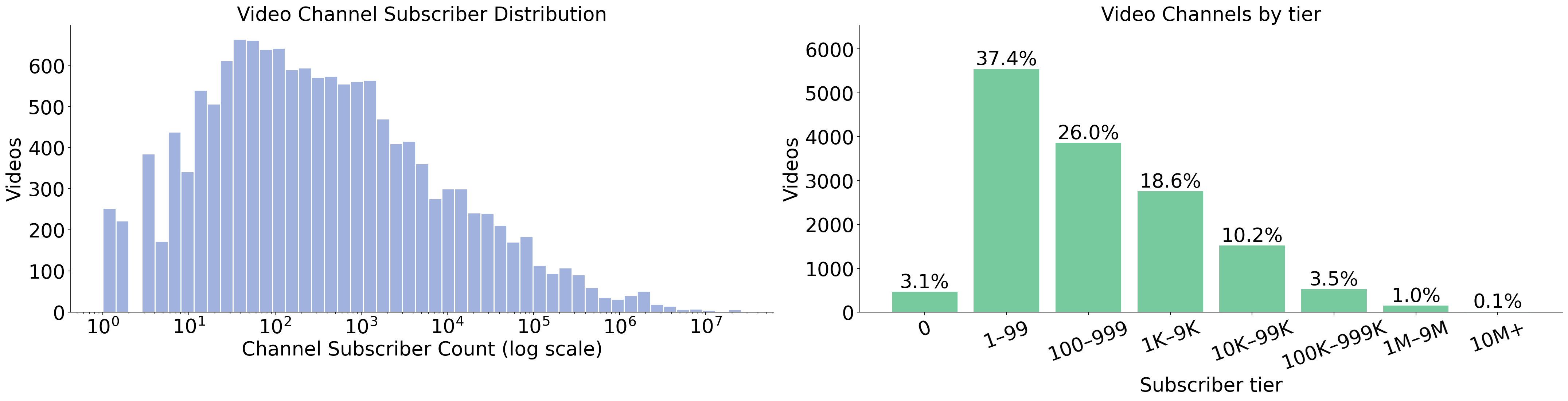}
\caption{Subscriber-count distribution of the source channels in \dataset. \textbf{Left}: histogram of subscriber counts on a log scale. \textbf{Right}: per-tier video count and dataset share (\%).}
\label{fig:subscriber_dist}
\end{figure}

Figure~\ref{fig:subscriber_dist} characterizes the subscriber-count distribution of the source channels behind the 14K unique videos in \dataset.
The histogram in the left panel spans subscriber counts from $1$ to $25.9$M on a logarithmic axis and peaks in the $1$K--$10$K range, while the right panel groups channels into eight tiers and reports each tier's share of the dataset.
Channels with fewer than $1$K subscribers account for approximately $67\%$ of all videos, whereas only $11$ videos ($0.07\%$) come from mega-channels with $10$M+ subscribers.
This heavy-tailed shape mirrors the long-tail creator ecosystem of YouTube as a whole, indicating that \dataset is not skewed toward top creators and supporting its representativeness as a substrate for studying micro-video popularity under realistic conditions.

\section{\proposed Details}
\label{sec:apn_b}

\subsection{Training}

\paragraph{Predictor head}
The regression head $\phi$ is a 2-layer MLP that maps the $<$REG$>$ token hidden state $h_i \in \mathbb{R}^{4096}$ to a scalar prediction:
$\text{Linear}(4096 \rightarrow 1024) \rightarrow \text{GELU} \rightarrow \text{Dropout}(0.1) \rightarrow \text{Linear}(1024 \rightarrow 1)$.
The head contains approximately 4.2M parameters, less than 0.06\% of the Qwen3-VL-8B-Instruct backbone.

\paragraph{Step~1: offline training}
We fine-tune $\{\theta, \phi\}$ jointly with AdamW (8-bit) for up to 7 epochs with early stopping (patience 2, monitoring validation MSE).
We use bf16 precision with tf32 enabled, SDPA attention, and gradient checkpointing.
The learning rate is $3\mathrm{e}{-5}$ with cosine schedule and warmup ratio $0.05$, weight decay $0.01$, gradient clipping $1.0$, and maximum sequence length $8192$.
Per-device batch size is $2$ with gradient accumulation $16$, giving an effective batch size of $64$ across two GPUs.
The loss balance coefficient is $\alpha{=}0.5$.
The rationale supervision targets are generated with GPT-4o-mini using the prompt template in Table~\ref{tab:rationale_prompt}, taking each video's multi-modal content $X_i$, evidence-card $E_i$, and ground-truth popularity tier $c_i$ as input.
This follows the distillation paradigm in which a teacher LLM first produces structured reasoning outputs that a task-specific model then learns to reproduce~\cite{kim2024self, seo2024make, leeimagine}.

\begin{table*}
\small
\centering
\caption{Prompt for generating dimension-wise rationales with saliency scores.}
\label{tab:rationale_prompt}
\setlength{\tabcolsep}{4pt}
\renewcommand{\arraystretch}{1.05}
\begin{tabular}{p{14cm}}
\toprule
\textbf{Evidence-aware rationale generation prompt} \\
\midrule
\textcolor{teal}{\textbf{[Task]}}\\
You are a video popularity analysis expert. Given a YouTube Shorts video's multimodal content, its open-web Evidence Card, and the coarse popularity tier of the video, generate an evidence-aware rationale for video popularity predictor training. Produce one line of reasoning per Evidence Card dimension (D1 Topic \& Entity Context, D2 Public Discourse, D3 Related Content Activity) explaining how that dimension contributes to the given popularity tier, and assign an integer saliency score from one to ten that measures the strength of its contribution. \\[2pt]
\textcolor{teal}{\textbf{[Inputs \& Constraints]}}\\
Use only the multimodal content (title; four key frames at zero, twenty-five, fifty, and seventy-five percent positions, where the zero percent frame is the thumbnail; and the ASR transcript split into segments temporally aligned with each frame), the Evidence Card (its D1, D2, and D3 evidence), and the coarse popularity tier. Do not use or mention channel identity, source region, source query, subscriber count or tier, follower count, the exact raw day-seven view count, or any log-transformed view count. The popularity tiers are: \textsc{Micro} (fewer than one thousand views), \textsc{Small} (one thousand to under ten thousand), \textsc{Medium} (ten thousand to under one hundred thousand), \textsc{Large} (one hundred thousand to under one million), and \textsc{Viral} (one million or more). Score saliency by how strongly the dimension explains the provided popularity tier (one to three for weak, four to seven for moderate, eight to ten for strong), not by how impressive the evidence sounds in isolation; do not output numeric view counts, do not assign a separate popularity tier to each dimension, and keep every explanation concise and grounded in the Evidence Card. \\[2pt]
\textcolor{teal}{\textbf{[User Prompt Template]}}\\
Title: \texttt{\{title\}}\quad Popularity tier: \texttt{\{popularity\_tier\}}\\
Interleaved frames and ASR (in temporal order):\\
\quad[Frame 0\% / thumbnail]: \texttt{\{frame\_0\}}\quad ASR (0--25\%): \texttt{\{asr\_seg\_0\}}\\
\quad[Frame 25\%]: \texttt{\{frame\_25\}}\quad ASR (25--50\%): \texttt{\{asr\_seg\_25\}}\\
\quad[Frame 50\%]: \texttt{\{frame\_50\}}\quad ASR (50--75\%): \texttt{\{asr\_seg\_50\}}\\
\quad[Frame 75\%]: \texttt{\{frame\_75\}}\quad ASR (75--100\%): \texttt{\{asr\_seg\_75\}}\\
D1 evidence: \texttt{\{d1\_evidence\}}\quad D2 evidence: \texttt{\{d2\_evidence\}}\quad D3 evidence: \texttt{\{d3\_evidence\}} \\[2pt]
\textcolor{teal}{\textbf{[Output Format \& Schema]}}\\
\texttt{D1/D2/D3: <reasoning> | Saliency: <1-10>} \\
\texttt{\{"dimension\_analysis": \{"D1": \{"reasoning": <str>, "saliency": <int 1-10>\}, "D2": \{...\}, "D3": \{...\}\}\}} \\
\bottomrule
\end{tabular}
\end{table*}

\paragraph{Step~2: online adaptation}
We attach LoRA adapters (rank $r{=}16$, scaling $\alpha_{\text{LoRA}}{=}32$) to all attention projections of the frozen backbone $\theta^{(1)}$ and update only $\{\Delta, \phi\}$.
We use AdamW with learning rate $1.2\mathrm{e}{-4}$ and per-update batch size $4$.
Adaptation is triggered every $M{=}32$ new drift samples and uses an adaptation set $\mathcal{M} = \mathcal{B}_D \cup \mathcal{B}_A$ with drift buffer capacity $N_D{=}256$ and anchor buffer $N_A{=}1024$.

\paragraph{Threshold calibration}
The growth-ratio thresholds $(\gamma_{\text{low}}, \gamma_{\text{high}})$ are calibrated on $\mathcal{D}_{\text{val}}$ so that the most extreme 15\% of growth ratios at each tail qualify as \textit{delayed viral} ($\gamma_i < \gamma_{\text{low}}$) and \textit{initial viral} ($\gamma_i > \gamma_{\text{high}}$), following the growth-curve taxonomy of~\cite{france2016characterizing}.
The error threshold $\delta_y{=}2.0$ corresponds to roughly the upper quartile of absolute prediction errors on $\mathcal{D}_{\text{val}}$, ensuring that only large errors are admitted as drift candidates.

\paragraph{Compute}
All experiments are conducted on $2\times$NVIDIA H200 (144GB) GPUs, totaling approximately $250$ GPU-hours including offline training, online adaptation, and preliminary experiments.

\subsection{Offline Baseline Implementations}

\paragraph{Video Content-Only}
We fine-tune the same Qwen3-VL-8B-Instruct backbone on each video's multi-modal content $X_i$ (title, ASR transcript, thumbnail, and sampled key frames) without any external context.
The variant with the \texttt{<REG>} regression head matches \proposed's prediction interface, while the variant without it generates the popularity score directly as text.

\paragraph{Video Corpus Retrieval-Augmented}
Following~\cite{zhong2024mmra, cheng2025seeing, cheng2024ragtrans, chen2025echoes, cheng2025incontext, xu2025skapp}, we encode the four sampled key frames of each video with the CLIP-ViT-B/32 vision encoder~\cite{radford2021learning} and its title and ASR transcript with AnglE-BERT~\cite{li2023angle}, using the concatenation as retrieval keys, and build an offline retrieval bank from $\mathcal{D}_{\text{train}}$ ($\sim$8K videos).
At inference, we retrieve the top-5 nearest historical neighbors per target video; each neighbor is serialized as ``video ID, title, channel, similarity score, ASR transcript, and training label $\log_2(v_i(7){+}1)$'' and appended to $X_i$ as additional context.

\paragraph{Open-Web Grounded — Raw Web Snippet}
We retrieve top-5 unstructured search results per target video through the Google Custom Search JSON API, using the video title as the query.
To prevent in-platform leakage, we exclude YouTube and its related domains (\texttt{youtube.com}, \texttt{youtu.be}, \texttt{m.youtube.com}) from both the query and the returned results.
Each result is serialized as ``title, snippet, URL'' and concatenated to $X_i$, providing the predictor with raw open-web context without the structured evidence-card design of \proposed.

\subsection{Online Baseline Implementations}

\paragraph{Cache-$k$NN}
Following~\cite{khandelwalgeneralization, bhardwaj-etal-2023-adaptation}, we maintain a FIFO cache of size $1000$ that stores \texttt{<REG>} hidden state, revealed label$)$ pairs from the stream and adjusts each prediction by interpolating with the mean label of the top-$k{=}8$ nearest cached neighbors under cosine similarity.

\paragraph{Temporal-Decay Residual Memory}
Following~\cite{zhang2023adanpc, lyu2026ts}, we maintain a per-topic running residual using exponential decay $\rho{=}0.9$, where the topic of each video is determined by its seed query (Section~\ref{sec:dataset}). Each prediction is corrected by the current per-topic residual estimate.

\paragraph{OGD-LoRA and ER-LoRA}
Following~\cite{wei2025onlinelora, wang2023orthogonal} for OGD-LoRA and~\cite{zhang2022a} for ER-LoRA, both baselines share \proposed's LoRA configuration ($r{=}16$, $\alpha_{\text{LoRA}}{=}32$, attention-only, lr $=1.2\mathrm{e}{-4}$) but update on every revealed batch without drift filtering. ER-LoRA additionally uses experience replay from a uniformly sampled buffer matched in size to $N_A$.

\subsection{Evidence-Card Effect Analysis}
\label{subsec:appendix_ec_details}

\paragraph{Effect of evidence-card dimensions}
We fix \proposed~(\textit{Offline}) and ablate one dimension of the evidence-card at a time on a 1{,}000-sample subset.
Table~\ref{tab:ec_dimension_ablation} shows that the three dimensions contribute unequally: removing D3 (Related Content Activity) causes the largest degradation across all metrics (nMSE $0.632 \rightarrow 0.716$, SRC $0.681 \rightarrow 0.509$), indicating that signals about competing and amplifying content most directly shape view-share outcomes by characterizing the attention space the target video must enter.
Removing D1 (Topic \& Entity Context) yields the second-largest drop (nMSE $\rightarrow 0.692$, SRC $\rightarrow 0.567$), confirming that factual grounding of the topic is essential for calibrated tier prediction.
By contrast, removing D2 (Public Discourse) leaves all metrics essentially unchanged and even slightly improves ranking (SRC $0.681 \rightarrow 0.686$), suggesting that discourse-level signals overlap substantially with the topic and competing-content context already captured by D1 and D3, and may introduce sentiment-level noise that dilutes the regression signal.
This indicates that competing-content saturation and topic grounding carry the bulk of the evidence-card's predictive value, while public-discourse refinement remains an open direction, partly because reliably parsing fine-grained reactions from diverse online content is inherently difficult even for modern LLMs~\cite{heo2025can}.

\begin{table}[h]
\caption{Per-dimension ablation of the evidence-card with \proposed~(\textit{Offline}) as the predictor.}
\vspace{0.2cm}
\label{tab:ec_dimension_ablation}
\centering
\small
\renewcommand{\arraystretch}{1.05}
\setlength{\tabcolsep}{6pt}
\begin{tabular}{lcccc}
\toprule
\textbf{Condition} & nMSE ($\downarrow$) & MAE ($\downarrow$) & SRC ($\uparrow$) & PCC ($\uparrow$) \\
\midrule
Full evidence-card                      & \textbf{0.632} & \textbf{2.324} & \underline{0.681} & \underline{0.668} \\
\;\; w/o D1 (Topic \& Entity)           & 0.692 & 2.586 & 0.567 & 0.606 \\
\;\; w/o D2 (Public Discourse)                 & \underline{0.636} & \underline{2.337} & \textbf{0.686} & \textbf{0.672} \\
\;\; w/o D3 (Related Content)           & 0.716 & 2.596 & 0.509 & 0.589 \\
\bottomrule
\end{tabular}
\end{table}

\paragraph{Effect of evidence-card alignment}
We fix \proposed~(\textit{Offline}) and replace each video's own evidence-card with one drawn from another video on a 1{,}000-sample subset, comparing three alignments: the video's \textit{own} evidence-card, a \textit{topic-matched} card from another video sharing the same seed query, and a \textit{random} card from an unrelated video.
Table~\ref{tab:ec_alignment} shows that misaligned evidence collapses prediction quality even when the substitute card shares the same topic: random alignment inflates nMSE from $0.911$ to $1.448$ and cuts SRC by more than half ($0.592 \rightarrow 0.237$), while topic-matched substitution recovers only a small fraction of this loss (nMSE $1.219$, SRC $0.313$).
This indicates that the predictive value of the evidence-card is not a topic-level prior shared across videos on the same theme, but a sample-specific signal whose alignment to the target video carries most of the information.
Together with the structuring effect in Table~\ref{tab:offline_main}, where unstructured raw web snippets degrade ranking below the content-only baseline, evidence-cards are useful only when they are both structured and aligned to the specific video being predicted.

\begin{table}[h]
\caption{Evidence-card alignment ablation with \proposed~(\textit{Offline}) as the predictor.} 
\vspace{0.2cm}
\label{tab:ec_alignment}
\centering
\small
\renewcommand{\arraystretch}{1.05}
\setlength{\tabcolsep}{6pt}
\begin{tabular}{lcccc}
\toprule
\textbf{Alignment} & nMSE ($\downarrow$) & MAE ($\downarrow$) & SRC ($\uparrow$) & PCC ($\uparrow$) \\
\midrule
Own (target's own EC)                 & \textbf{0.911} & \textbf{2.856} & \textbf{0.592} & \textbf{0.547} \\
Topic-matched (same seed query)       & \underline{1.219} & \underline{3.304} & \underline{0.313} & \underline{0.242} \\
Random (unrelated video)              & 1.448 & 3.655 & 0.237 & 0.189 \\
\bottomrule
\end{tabular}
\end{table}

\subsection{Qualitative Explainability}
\label{subsec:appendix_explainability}

\begin{figure}[t]
\centering
\includegraphics[width=0.99\linewidth]{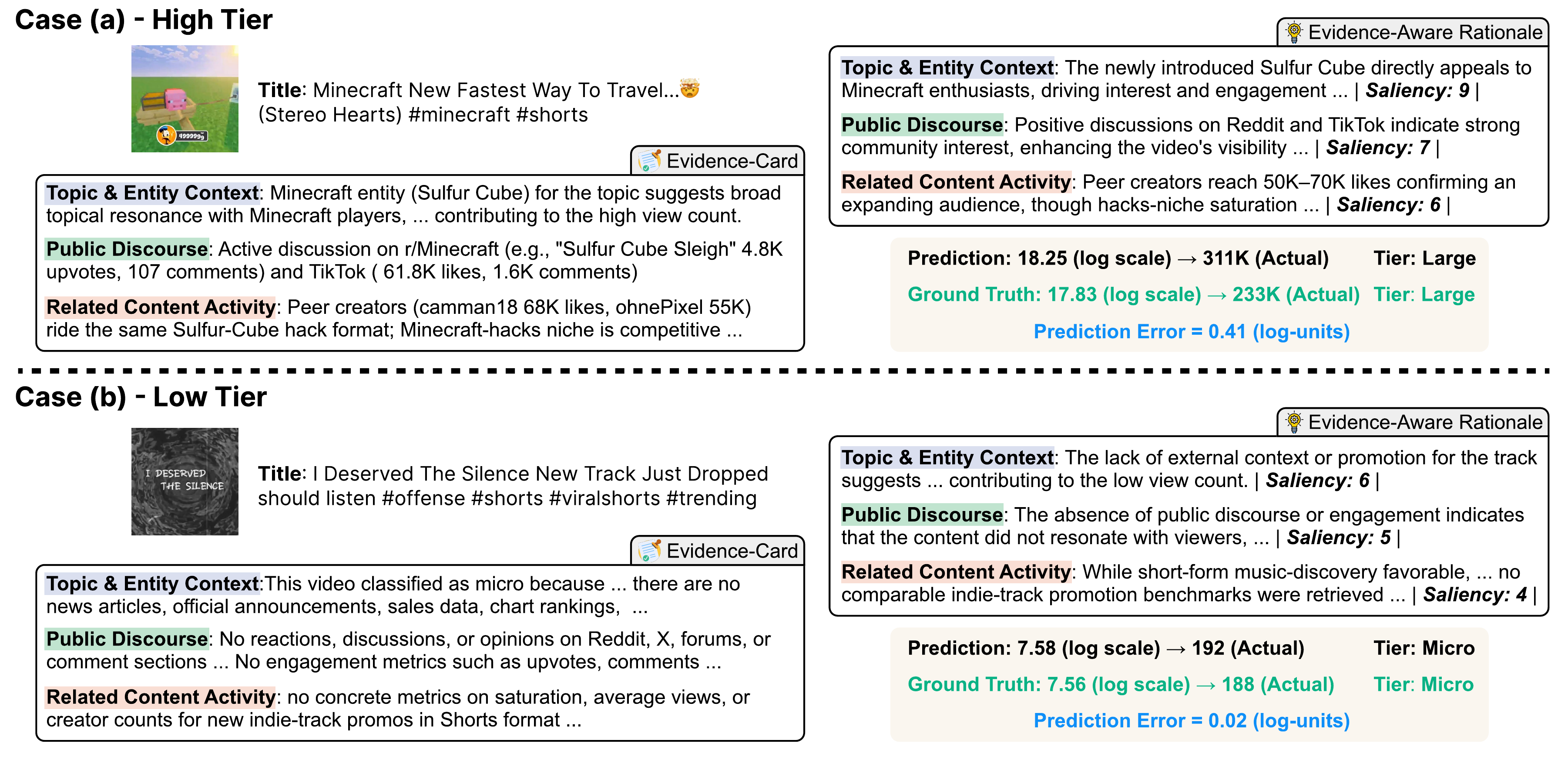}
\caption{Case study of \proposed on a Large-tier video (a) and a Micro-tier video (b), pairing the input evidence-card (left) with the generated rationale and per-dimension saliency (right).}
\label{fig:case_study_explain}
\end{figure}

Figure~\ref{fig:case_study_explain} shows \proposed on a Large-tier Minecraft Shorts (a), where the rationale grounds its prediction in three concrete evidence-card signals --- the ``Sulfur Cube'' entity, Reddit/TikTok discourse at $4.8$K upvotes and $61.8$K likes, and peer creators reaching $50$--$70$K likes --- and assigns saliency $9/7/6$ to D1/D2/D3, predicting $18.25$ against ground-truth $17.83$ (error $0.41$ log-units, both Large).
On the Micro-tier indie-track upload (b), the rationale conditions on the symmetric absence of these signals: no news coverage in D1, no discourse engagement in D2, and no comparable promotion benchmark in D3, yielding saliency $6/5/4$ and predicting $7.58$ against $7.56$ (error $0.02$, both Micro).
The same rationale-conditioned mechanism thus produces accurate predictions at both ends of the popularity distribution, with each prediction traceable to the specific evidence-card phrases the regression head conditioned on.
Together with the per-dimension saliency, this provides an evidence-grounded explanation of each prediction as an inherent component of \proposed's output, rather than as a post-hoc interpretation produced separately from the model.

\end{document}